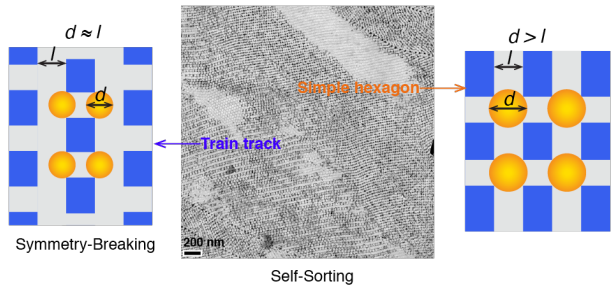

Symmetry-Breaking    Self-Sorting

$d \approx l$    Train track    Simple hexagon    $d > l$

# Symmetry-Breaking and Self-Sorting in Block Copolymer-based Multicomponent Nanocomposites

Le Ma[1,2]†, Hejin Huang[3]†, Peter Ercius[4], Alfredo Alexander-Katz[3] and Ting Xu[1,2,5]*

[1]Department of Materials Science and Engineering, University of California, Berkeley, CA 94720, USA

[2]Materials Sciences Division, Lawrence Berkeley National Laboratory, Berkeley, CA 94720, USA

[3]Department of Materials Science and Engineering, Massachusetts Institute of Technology, Cambridge, MA, 02139, USA

[4]National Center for Electron Microscopy, Molecular Foundry, Lawrence Berkeley National Laboratory, Berkeley, CA 94720, USA

[5]Department of Chemistry, University of California, Berkeley, CA 94720, USA

† Authors contributed equally

*Corresponding author email: tingxu@berkeley.edu

**Abstract**

Co-assembly of inorganic nanoparticles (NPs) and nanostructured polymer matrix represents an intricate interplay of enthalpic or entropic forces. Particle size largely affects the phase behavior of the nanocomposite. Theoretical studies indicate that new morphologies would emerge when the particles become comparable to the soft matrix's size, but this has rarely been supported experimentally. By designing a multicomponent blend composed of NPs, block copolymer-based supramolecules, and small molecules, a 3-D ordered lattice beyond the native BCP's morphology was recently reported when the particle is larger than the microdomain of BCP. The blend can accommodate various formulation variables. In this contribution, when the particle size equals the microdomain size, a symmetry-broken phase appears in a narrow range of particle sizes and compositions, which we named the "train track" structure. In this phase, the NPs aligned into a 3-D hexagonal lattice and packed asymmetrically along the $c$ axis, making the projection of the $ac$ and the $bc$ plane resemble train tracks. Computation studies show that the broken symmetry reduces the polymer chain deformation and stabilizes the metastable hexagonally perforated lamellar morphology. Given the mobility of the multicomponent blend, the system shows a self-sorting behavior: segregating into two macroscopic phases with different nanostructures based on only a few nanometers NP size differences. Smaller NPs form "train track" morphology, while larger NPs form "simple hexagon" structure, where the NPs take a symmetric hexagonal arrangement. Detailed structural evolution and simulation studies confirm the systematic-wide cooperativity across different components, indicating the strong self-regulation of the multicomponent system.

Hierarchical design and assembly of building blocks across different length scales is the essential characteristic of many functional materials and the living organisms that evolved in nature.[1-5] Significant efforts have been devoted to investigating/monitoring the interplay between components with different characteristic sizes to achieve desired structures and explore optimal performances.[6-10] Organic/inorganic blends are ideal systems for understanding the interplay between their constituents since they combine the compressible organic components and incompressible inorganic fillers where each individual with adjustable size.[11-14] One particular challenge is the implementation of co-assembly of each component instead of macroscopic phase separation *via* tuning the interactions, especially when the sizes of the building block are comparable.[15]

Co-assembly of inorganic nanoparticles (NPs) and nanostructured soft matrix represent a balance among intricate interplay of enthalpic or entropic forces directed toward low free energy morphology.[11, 16] Block copolymer (BCP) is one of the most well-studied nanostructured soft matrix and it has proven to be effective for directing the controllable assembly of NPs into hierarchical structures.[11-12, 17-21] The phase behavior of the BCPs/NPs blend is closely related to the relative length scale between the NP size and the microdomain of the soft matrix. When the particle size is small (smaller than the microdomain size of BCPs), the morphology is templated by the framework of the polymer matrix, where the enthalpic interactions between NPs and host microdomain dominate the assembly process.[18] The assembly usually follows the equilibrium structures of BCPs, and it has been challenging to fabricate structures beyond BCP's native morphologies. Theoretical studies show the potential to fabricate unconventional structures when the particle size is comparable or larger than the characteristic size of the polymer matrix. For example, researchers[22] observed a new self-assembled morphology distinctive from the original

BCP when the NP size is comparable to the radius gyration of the minority block, where the particles assemble inside the copolymer micelles. The resulting phase is attributed to the interplay between the particle-particle excluded-volume interactions, preferential particle/block interactions, and the entropic interactions related to polymer stretching. However, this concept has rarely been supported experimentally; NPs comparable to microdomain size prefer to stabilize the defects and form macroscopic phase separation between the particle-aggregation and pure BCP phases.[15] Because the favorable enthalpic interactions between NPs and the host block cannot overwhelm the significant loss in conformational entropy associated with polymer chain stretching around these large obstacles, which prevents us from investigating the interplays between components and exploring unique co-assembled morphologies.

Our recent study[23] proposed an entropy-driven approach that enables the NPs larger than the polymer microdomains to co-assemble with other components to form an ordered 3-D structure. This co-assembly behavior is achieved by a multicomponent blend composed of NPs, BCP-based supramolecules, and small molecules. The distribution of small molecules mediates unfavorable interactions and allows ordered lattice formation with high structural fidelity while accommodating various formation parameters. This multicomponent blend approach provides a platform to study the interactions among building blocks across different length scales, especially when a particle is similar or larger than the polymer microdomain, opening up possibilities to explore novel morphologies.

Herein, in the same blend, when the particle size equals to the microdomain size, we obtained a symmetry-breaking phase named the "train track" structure that occurs in a very narrow range of particle size and composition, despite the formulation flexibility provided by the multicomponent blends. Symmetry breaking has been shown a common way in 2D layered materials to introduce new properties[24-26] but hasn't been well-explored in soft matter. Here, NPs organized into a 3-D hexagonal lattice and packed asymmetrically along the *c* axis, leading to projections at *ac* and *bc* planes resemble train tracks. The broken symmetry of NPs reduced the polymer chain deformation and stabilized the metastable HPL morphology, and it can only take place in a very narrow window of particle sizes; Because the particle should be large enough to induce the polymer morphology to transform into HPL, but it has to be small enough to diffuse through the channel that connects the two adjacent layers of HPL to achieve structural stability. The narrow window of the "train track" structure also brings in additional control of the phase behavior of the multicomponent system. NP size with a 10% difference will lead to a different morphology. Considering the NPs used in the blend are not perfectly monodispersed (Figure S15), we observed another "simple hexagon" morphology in the same blend, wherein the NP size is a bit larger than that in the "train track" structure. Thus, the blend self-sorting into two macroscopic phases with different nanostructures based on a few nanometer differences in NP size, reflecting the precise control over interplay among building blocks and suggesting the strong self-regulation of the multicomponent system. We analyzed the detailed 2-D and 3-D morphologies formed in the blends, performed the molecular dynamic simulation to illustrate the spatial distribution of each component and explain the mechanism of structure formation, monitored the structure evolution and the kinetic pathway of the assembly process to further confirm the underlying basis of the unique phase behavior. Besides experimentally generating the unique symmetry-breaking morphology in organic/inorganic blends, the present study reflects the interplays of components in the multicomponent system when particle size is the same as the microdomain of the polymer matrix, and shows the self-regulation and the hierarchical structural control abilities of the multicomponent nanocomposite.

**Results and Discussion**

Specifically, the supramolecule, "PS(33 kDa)-b-P4VP(8 kDa)(PDP)$_1$", is constructed from 3-pentadecylphenol (PDP) hydrogen-bonding to the pyridyl side chains in polystyrene-block-poly(4-vinyl pyridine) (PS-b-P4VP) at a 1:1 molar ratio.[27] Free small molecules PDP (ratio to P4VP is 1:1) were added to the system to tune each microdomain's effective volume fraction. More importantly, it can modulate the interactions between components, as its solubility parameter is between those of two microdomains in the supramolecule.[28-29] NPs are alkyl-passivated iron oxide with size (15 nm) similar to the microdomain size of the supramolecule (See Figure S15 for NP size distribution). NP loading is 5 *vol%* if not specified.

**Morphology Characterization**. We obtained two ordered assemblies in the blend: one we named the "train track" structure, and the other we named the "simple hexagon" structure (Figure S1). We performed transmission electron microscopy (TEM) and high-angle annular dark-field scanning transmission electron microscopy (HAADF-STEM) tomography to characterize their 2-D and 3-D structures. Figures 1a and 1b show the TEM images of the "train track" structure. In this projection, NPs arrange into multiple sets of two parallel long chains, which look similar to the train tracks. There are gaps between NPs within one chain, and NPs from different chains are aligned with each other. STEM tomography reconstruction shows a hexagonal lattice perpendicular to the long axis of the "train track" projection (Figure 1c, 1d, and Movie S1). It would represent the first time to our best knowledge that the particles self-assemble into an asymmetric 3-D ordered structure in polymer nanocomposites. The morphology remains unchanged after thermal annealing at 110 °C overnight, indicating the structure is relatively thermodynamic stable (Figure S2). However, this structure only takes place in a narrow range of particle size and blends compositions: NPs form chain-like morphology if there are no free small molecules in the system (Figure S3); The blend adopts different morphologies if the NP size is smaller or larger than the host microdomain size. With the same composition, the blend with smaller particles forms cylinder morphology,[30] while the composite with larger particles forms the "simple hexagon" structure. Detailed discussion regarding the NP size will be presented later. Figures 1e and 1f show the TEM images of the other structure we observed in the blend. NPs form highly ordered arrays with a "square-like" lattice in the presented projection. Based on the STEM tomography reconstruction, the 3-D structure was determined to be square lattice out-of-plane and hexagon lattice in-plane (Movie S2), which we named the "simple hexagon" structure.

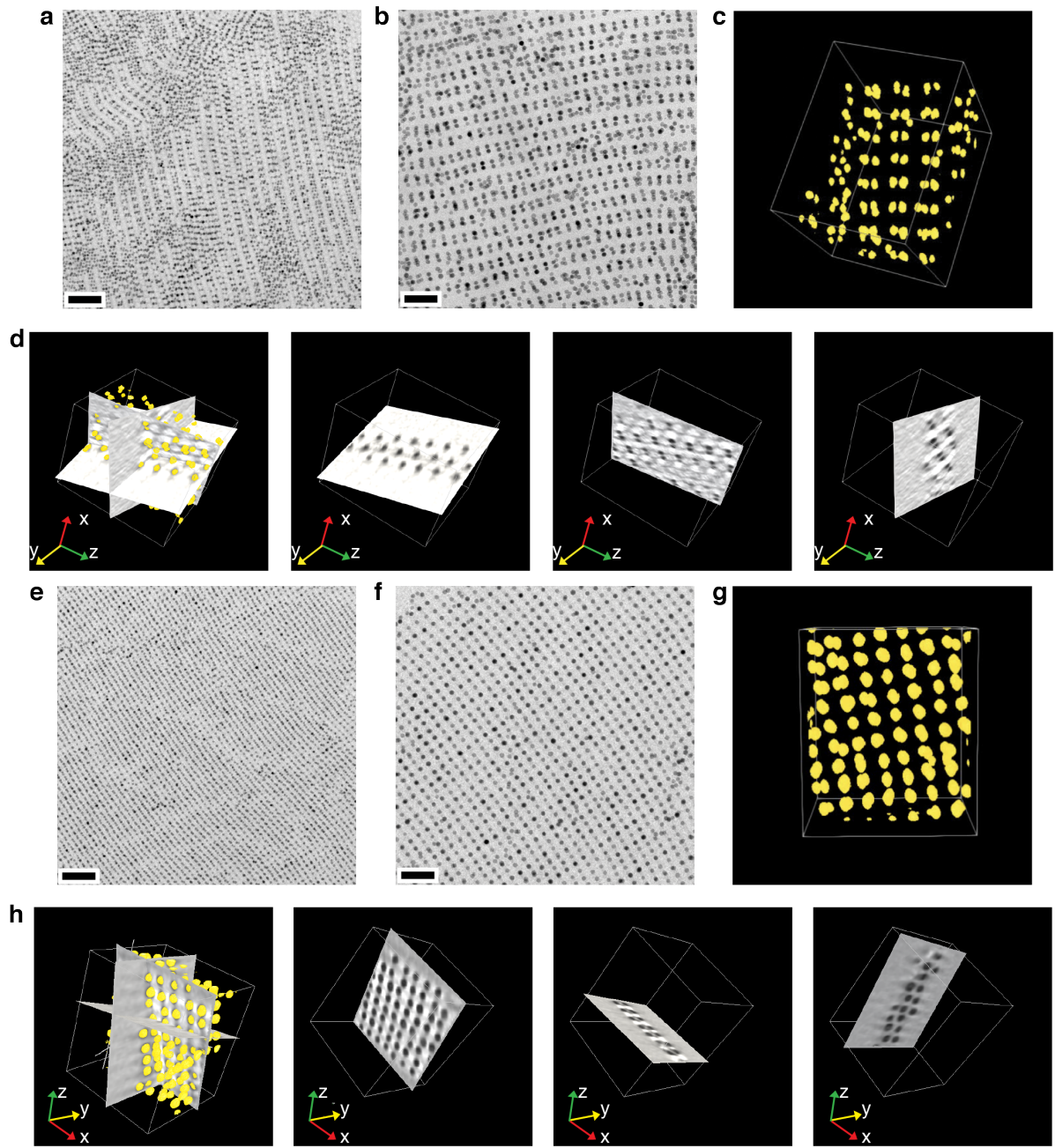

**Figure 1. Two ordered assemblies ("train track" and "simple hexagon") observed in the multicomponent nanocomposite. a-d,** Morphology characterization of the "train track" structure. **a, b,** TEM images. **c**, 3-D tomography. **d**, 3 slices of the reconstruction results. The yellow dots in the first image are NPs. The black dots in the last three images are NPs. **e-h,** Morphology characterization of the "simple hexagon" structure. **e, f,** TEM images. **g,** 3D-tomography. **h,** 3

slices of the reconstruction results; The yellow dots in the first image are NPs. The black dots in the last three images are NPs. Scale bars: **a, e,** 200 nm. **b**, **f,** 100 nm.

**Self-Sorting Behavior of the Multicomponent System.** The occurrence of two different morphologies in the same blend is primarily related to the NP size. Analysis of the NP size distribution in the two structures within the same sample (Figure 2a) shows that the average size of NP is slightly smaller in the "train track" structure than that in the "simple hexagon" structure (Figure 2b and 2c). The presence of macro-grains with different nanostructures indicating a self-sorting behavior of the multicomponent system. Mixtures of NPs with different sizes can spontaneously segregate and assemble into two morphologies: smaller NPs prefer to form the "train track" morphology while the relatively large NPs assemble into the "simple hexagon" structure. The geometric incompatibility between the nanostructures then drives macroscopic phase separation of the blend, leading to the formation of "train track" phase and the "simple hexagon" phase (Figure 2a). Small-angle X-ray scattering (SAXS) also confirmed the existence of the two structures in a relatively large range. Figure 2d shows the 2-D scattering pattern of the assembled morphologies. The inhomogeneous intensity distribution around the diffraction rings is due to the surface-induced orientations, as the sample solution was dried within a small tube. The fitted pattern suggests there is a combination of the "train track" and "simple hexagon" structures as shown in Figures 2e and 2f (Figure S4). For "train track" morphology, a = b≠ c and $\alpha = \beta = 90°$, $\gamma = 120°$. There is another particle in the 1/3 of the c-axis, leading to an asymmetric projection at *ac* and *bc* plane (Figure. 2f). The "train track" structure belong to the P6mm space group. For "simple hexagon" morphology, a = b≠ c and $\alpha = \beta = 90°$, $\gamma = 120°$, belong to the P6/mmm space group. The characteristic peaks for "train track" and "simple hexagon" are labeled by blue and orange bars, respectively. For example, $q$ ~0.024 Å$^{-1}$ is the (101) of "simple hexagon", and $q$ ~0.029 Å$^{-1}$ is the (111) of "train track".

We believe the segregation based on only a few nanometers difference indicates a strong self-regulation behavior of our multicomponent system and also the long-range cooperation between different components. It also suggests the reduced kinetic barrier across different interfaces as the distribution of free small molecules.[31] As a result, the components can easily rearrange based on the local needs, such as NP size, to achieve the self-sorting behavior. The resulting macroscopic phase separation with internally ordered assembly in each phase demonstrates that we can access structure control over multiple length scales.

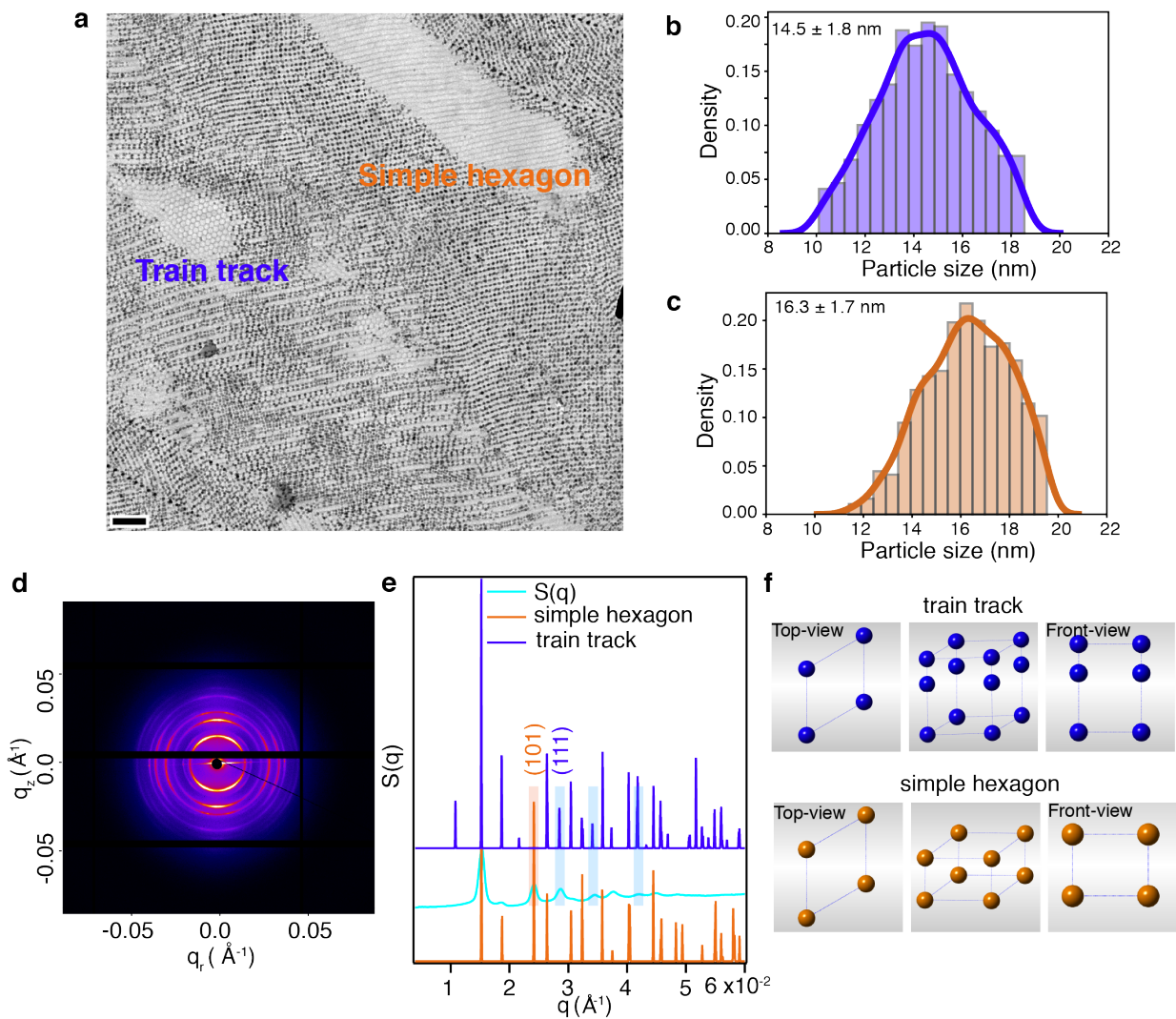

**Figure 2. Self-sorting behavior of the multicomponent nanocomposite**. **a**, TEM image of the blends with the coexist of two structures. The lower left region shows the "train track" structure, and the upper right region shows the "simple hexagon" structure. Scale bar: 200 nm. **b, c,** NP size distribution in the two morphologies. **d, e,** 2-D SAXS pattern and the fitted profiles of the multicomponent nanocomposite. S(q) is the structure factor of the blend. The "train track" and "simple hexagon" curves are calculated diffraction patterns. **f**, the unit cell of "train track" structure and "simple hexagon" structure. The lattice parameters for "train track" structure are: $a = b = 47.6\ nm,\ c = 58\ nm,\ \alpha = \beta = 90°,\ \gamma = 120°$. There is another NP at ~1/3 of the $c$ axis. The lattice parameters for "simple hexagon" are: $a = b = 47.6\ nm,\ c = 33.5\ nm,\ \alpha = \beta = 90°,\ \gamma = 120°$.

**Dissipative Particle Dynamics (DPD) Simulation Elucidates the Structural Formation Mechanism.** DPD simulation was carried out to explain the underlying mechanism of the new morphology and the self-sorting behavior of the multicomponent blends. The studied 4-component blend contained BCPs (with a coil-comb structure to mimic the supramolecules PS-b-P4VP(PDP)$_1$), free small molecules, and NPs (see details in Materials and Methods). Simulation results in Figure 3a reveal the assembled morphology and each component's distribution. In the "train track" structure, supramolecule adopts hexagonally perforated layer (HPL) morphology that consists of alternating layers of the PS-rich domain and P4VP-rich domain. The PS-rich layer is the minority component and contains hexagonally packed cylindrical perforations of the majority P4VP-rich block, where the holes across adjacent layers are out of registry (Figure S5). NPs are located in every other HPL layer and are aligned within cylinders, thus forming an asymmetric ordering. The morphology in projection parallel to the perforated layer (*yz* plane) is consistent with the second slice of the tomography reconstruction in Figure 1d. In the projection normal to the perforated layers (*xz* plane), NPs arrange into hexagonal lattice. This 3-D arrangement can be held over one million dynamic steps after releasing the NPs (Movie S3), further confirming the structure's stability. Unlike the "train track" structure, NPs assemble into a symmetrical arrangement in the "simple hexagon" structure, and polymers form a perforated layer morphology consisting of alternating layers of the PS-rich domain and P4VP-rich domain with holes aligned across adjacent layers (Figure 3a).

Without NPs incorporation, supramolecule/PDP blend adopts a hexagonally packed cylindrical morphology constituted by a PS cylinder surrounded by P4VP(PDP) domain. The two ordered assemblies in the multicomponent nanocomposite indicate the addition of NPs induces the morphological transition and stabilizes the perforated layer structure. Because the nanoparticle size is close to the microdomain size, addition of NPs would bring undulation along with the cylinder and makes the cylindrical morphology unstable. Small molecules will redistribute themselves to mediate the unfavorable interactions, allow the incorporation of large particles and minimize the system's free energy, which induces the effective volume fraction change of the PS-rich and P4VP-rich microdomains, resulting in the morphology transition. This is different from previous studies where the large NP prefer to stabilize the defects of BCPs and form macrophase separation. Here, the multicomponent system self-regulates the spatial distribution of its components, and allows generating new morphology absent in traditional BCP/NP blends. Based on the simulation results, there are 16.65% of free small molecules in the PS-rich domain of the "train track" structure, while 10.42% of the "simple hexagon" structure (Figure S6).

Further simulation studies reveal that the train-track structure results from an order-order transition from the 'simple hexagon' structure when the particle size is similar to the microdomain size. This structural transformation from "simple hexagon" to 'train-track' depends on entropy-enthalpy compensation, which is directly affected by the size of NPs. This result also explains the underlying basic of the experimentally observed size-dependent assemblies. Figures 3b and 3c illustrate our hypothesis for the structure formation mechanism. During the self-assembly process, the multicomponent nanocomposite self-organizes to form the "simple hexagon" structure. This morphology, however, induces extra entropic penalty due to polymer chain stretching, as the holes of adjacent HPL layers prefer to be out of registry.[32] If the NP is larger than the microdomain size, each NP spans two adjacent PS-rich layers (Figure 3b). The enthalpic gain to have the holes in registry outweighs the entropic penalty of chain stretching, making the simple hexagon stable. If the size of nanoparticles is comparable to the microdomain size, however, NPs are not able to overcome the barrier to align the holes of every layer in the perforated layer (PL) phase. Furthermore, a Pierls-like instability takes place since having two aligned particles close to each other reduces the free energy of the system due to two factors: i) the chain stretching due to the formation of holes in the lamella has two minima for small particles, and ii) the small molecule prefers to be localized in the P4VP phase to reduce the enthalpic costs. The first factor is related to the arrangement of the chains in the holes of the PL phase. Previous studies have shown that a NP of small size will encounter two minima, one on each side of the middle of the hole, while larger particles will reside in the middle.[33-34] The second factor is straightforward and relates to the accumulation of small molecules in the "links" between different particles in different layers. Dimerization has been observed in electronic systems, but to the best of our knowledge it had not been observed in soft matter systems. Critical to this phenomenon is the lowering of the free energy upon breaking the 1-D symmetry. As mentioned above, this is due to the fact that particles need to select one of the two minima on each side of the lamella. Once it is selected, the opposite particle in the next layer must align to form a bridge. This simultaneously lowers the enthalpy and maximizes the entropy of the BCP chains.

DPD simulation using our reparametrized force field that captures bulk and thin film experimental morphologies[35] verified this hypothesis as shown in Figure 3. In Figure. 3a, we show the formation of the train track structure from disordered conditions. However, to prove this is the lowest state, we also considered the possibility when the NPs are first fixed into the "simple hexagon" structure (Figure 3c), then their positions in the y-direction are released to allow oscillations of NPs. The equilibrium morphology of the blend is related to the ratio between the NP size ($d$) and the polymer microdomain size ($l$), that is $d/l$. As shown in Figure 3d, where the $d/l$

≈1.0, after a relaxation period of one million time-steps, NPs transform into the "train track" structure (Movie S5a). This transition is consistent with our hypothesis and indicates the "train track" structure is more entropically favorable at this condition. In addition, the transition process is barrier-less and the neighbor distance for particle change continuously along one column (Figure S7). If the NP size is larger than the microdomain size, as shown in Figure 3e ($d/l \approx 1.1$), NPs always maintain a "simple hexagon" morphology even with the oscillation (Movie S5b). Considering the transition is an overall slow process as nucleation of the new phase requires a large amount of NPs in the adjacent layers to oscillate towards each other, the evaporation rate during the assembly process would affect the final morphology. Indeed, our control experiment with a fast evaporation rate shows more "simple hexagon" morphology than the slow drying case (Figure S8), further confirming that part of the "train track" structure is transformed from the "simple hexagon" structure. Future studies will be conducted to investigate the interface between the "simple hexagon" and the "train track" structures (Figure S9). Based on the previous size distribution analysis, the "train track" structures comprised of 14-16 nm NPs could be transformed from the "simple hexagon" structure, as most of the overlapped NP size between those two structures lies in this range. While the smaller NPs (< 14 nm) directly stabilize the "train track" structure and the larger ones (> 16 nm) always maintain the "simple hexagon" morphology. These results also confirm the self-sorting behavior of the system based on several nanometer differences in NP size.

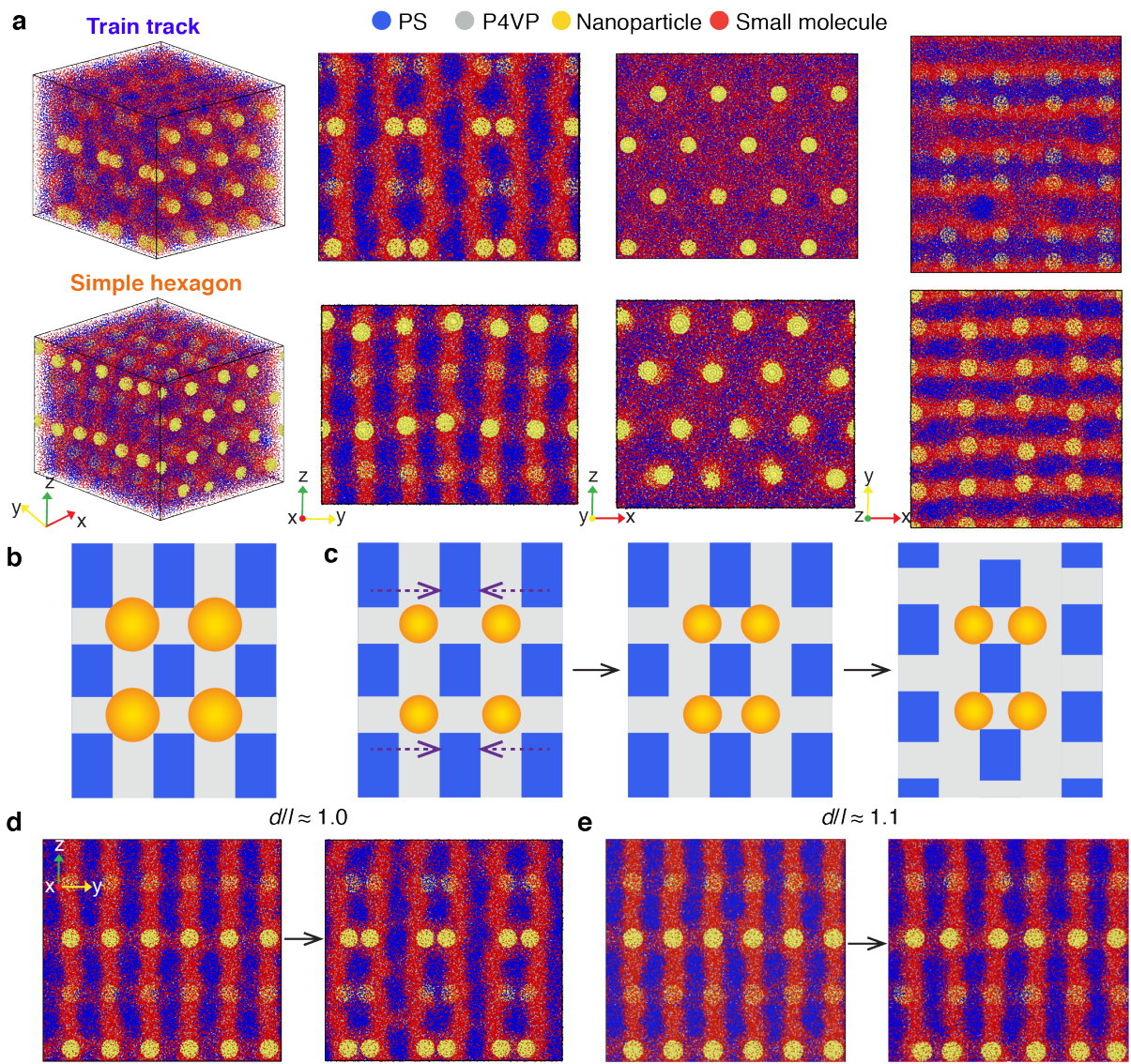

**Figure 3. Spatial distribution of each component in two ordered assemblies and their structure formation mechanism. a,** DPD simulation results for the "train track" structure and the "simple hexagon" structures. These images show the morphology and component distribution of the blend at different projections. All the parameters used for "simple hexagon" structure are the same as the "train track" simulation, except for the NP size, which is 10% larger than the "train track" structure. **b,** Schematic of the "simple hexagon" structure when the NP size is large. **c,** Schematic of the "simple hexagon" structure to "train track" structure transition. The purple arrows

show the NPs' oscillation. **d, e,** DPD simulation results for blends of coil-comb BCPs, small molecules, and NPs with different NP sizes. The NPs are first fixed into the "simple hexagon" structure (left image), then their positions in the *y*-direction are released to allow oscillations of NPs. The right images show the morphologies after a relaxation period of one million time steps. All the simulation setup parameters between d and e are the same except NP size. The size of NP in e is 10% larger than that in d.

**Structural Evolution**. To further understand the assembly pathway and confirm the structural formation mechanism of the two morphologies in the multicomponent nanocomposite, *in-situ* grazing-incidence transmission small-angle X-ray scattering (GTSAXS) was performed to investigate the assemble process during solvent evaporation. The nanocomposite solution was drop cast on a silicon wafer and the sample was monitored every 30 seconds during the drying process. We intentionally chose the initial film thickness to be ~2μm, which is ~70 times the periodicity of the supramolecule. Therefore, we can neglect the film's confinement effect and treat it as a bulk sample. The surface can help us to align the domains for easier analysis but not induce any morphological changes. GTSAXS is a relatively new technique that can be used to probe the interior of the film.[36-37] Based on our film thickness and the incident angle (0.8°), the distance that the beam travels through the sample is on the order of tens of microns during the measurement. Therefore, the scattering signal should mainly come from the film inside rather than the surface, which can provide insights into the structural evolution of the bulk sample's phase behavior. The complete set of the GTSAXS patterns is included in the supporting information (Movie S4 and Figure S10). Figure 4a shows the selected 2-D GTSAXS patterns. We observed three stages in the process of structural evolution: the ordered assemblies' formation, order-to-order transition, and local rearrangement.

In stage 1, clear diffraction spots emerge at a short drying time (3.5 min). These sharp diffraction spots indicate that the arrangement of NPs is a combination of a "train track" structure and a "simple hexagon" structure with a high degree of orientation. The highly ordered morphologies were driven by both interfaces, the substrate interface and the air-film interface. The rapid formation of the ordered assemblies suggests that the system has high mobility, which is related to the high solvent fraction at the early stage of the drying process. The estimated solvent fraction based on the film thickness is ~75 vol%. However, a high solvent content often leads to a weak enthalpic driving force for the assembly process since the effective Flory-Huggins segmental interaction parameter inversely depends on the solvent fraction.[38] The occurrence of the ordered structures at high solvent fractions indicates that entropy plays an essential role in the assembly process.[31]

In stage 2, there is an order-to-order transition with solvent evaporation, as indicated by the disappearance and appearance of the diffraction spots. At ~10 minutes, the diffraction peaks smeared out, and changed into a ring several minutes later. This observation shows that the ordered arrangement of NPs was disrupted. However, the ring indicates a ~30 nm interparticle correlation, which is related to the inherent correlation length of supramolecules. After ~10 mins (27 min), the clear diffraction peaks appear again, suggesting the system assemble into the ordered structure again. After the transition, there are still two ordered assemblies in the system, indicated by the similar structural peaks before and after the transition (see the scattering curves for 4 min and 27 min in Figure 4b). In Figure 4b, the (101) is the characteristic peak of the "simple hexagon" and (111) is that of the "train track". This observation provides evidence for the NPs induced morphological transition: the polymer matrix transformed from cylinder to perforated layer structure upon NPs incorporation, as shown in Figures S11. In the early stage of structure formation, the polymers assembled into hexagonal or square stacked cylinders parallel to the substrate, and NPs occupy the four corners of the interstitial sites. Previous studies show that NPs can act as fillers, localizing in the interstitial regions between cylindrical microdomains to effectively release the polymer chain deformation.[39] However, the NPs used here are larger than the interstitial size, which may lead to more chain deformation (Figure S12). Therefore, the blends self-regulate the distribution of each component to arrange in ways that minimize the free energy of the system. Polymers can arrange into hexagonal-packed cylinders with smaller NPs occupy four corners of the hexagon rather than six corners due to the large elastic strain of polymers induced by steric hindrance (Figure S12). Studies with a high particle loading confirmed that it is more energetic favorable for NPs to adopt the current arrangement occupying four corners of the hexagon. Both the experiment and simulation results show large distortions of hexagon if six corners are occupied (Figure S13). Polymers can also arrange into square-packed cylinders to increase the interstitial size, allowing the incorporation of relatively larger NPs (Figures S11 and S12). In those ways, the presence of NPs in the interstitial site can relieve the polymer chain stretching. In the meantime, the existence of NPs induces undulation along the length of cylinders,[40] leading to the formation of the modulated hexagonally packed cylinder.[41] The undulating interfaces approach each other and coalesce with solvent evaporation. Then the channels between adjacent cylinders formed and grew along the cylinders' axes and finally transformed into lamellar morphology,[41] as indicated by the boxes shown in Figures S11. Considering that the composition of the supramolecule/PDP blend is at the boundary of the cylinder and lamellae, and the perforated layer morphology appears as a metastable phase between these two morphologies,[42] the incorporation of nanoparticles will induce composition fluctuation and lead to the morphology transition, which actually stabilized the perorated layer morphology.

As part of the small molecules migrated to the PS-rich domain, as shown in previous simulation results, the effective volume fraction of P4VP(PDP)$_1$-rich phase decreases, leading to the cylinder to perorated layer transition. The hexagonal packed cylinders transform into HPL morphology with the holes from adjacent layer out of registry, while the square-packed cylinders transform into perforated layer structure with holes in different layers aligned with each other. During this order-to-order transition, the spatial arrangement of NPs remains unchanged, as shown in Figures S11, which explains the similar scattering peaks before and after transition in GTSAXS results. The transition can be further confirmed by TEM images that captured the intermediate state of structure evolution (Figure S14). As the solvent evaporation, some "simple hexagon" structures transformed into "train track" structures, indicated by the intensity changes of the two peaks marked in Figure 4c where the intensity of the (101) of "simple hexagon" decreases and the (111) of "train track" increases. This is consistent with the simulation results and provides the direct experimental evidences.

In stage 3, the system can only undergo local arrangement, leading to the broadening of the scattering peaks. The overall structure is set after stage 2. Once that structure has been formed, the system is unable to rearrange significantly and can only undergo short-range arrangement considering the low solvent fraction. Together, the *in-situ* GTSAXS study presents us with a pathway of how the system assembled into the two ordered structures, shows the NP induced order-to-order transition, captures "simple hexagon" to "train track" structure transition, confirming the structure formation mechanism.

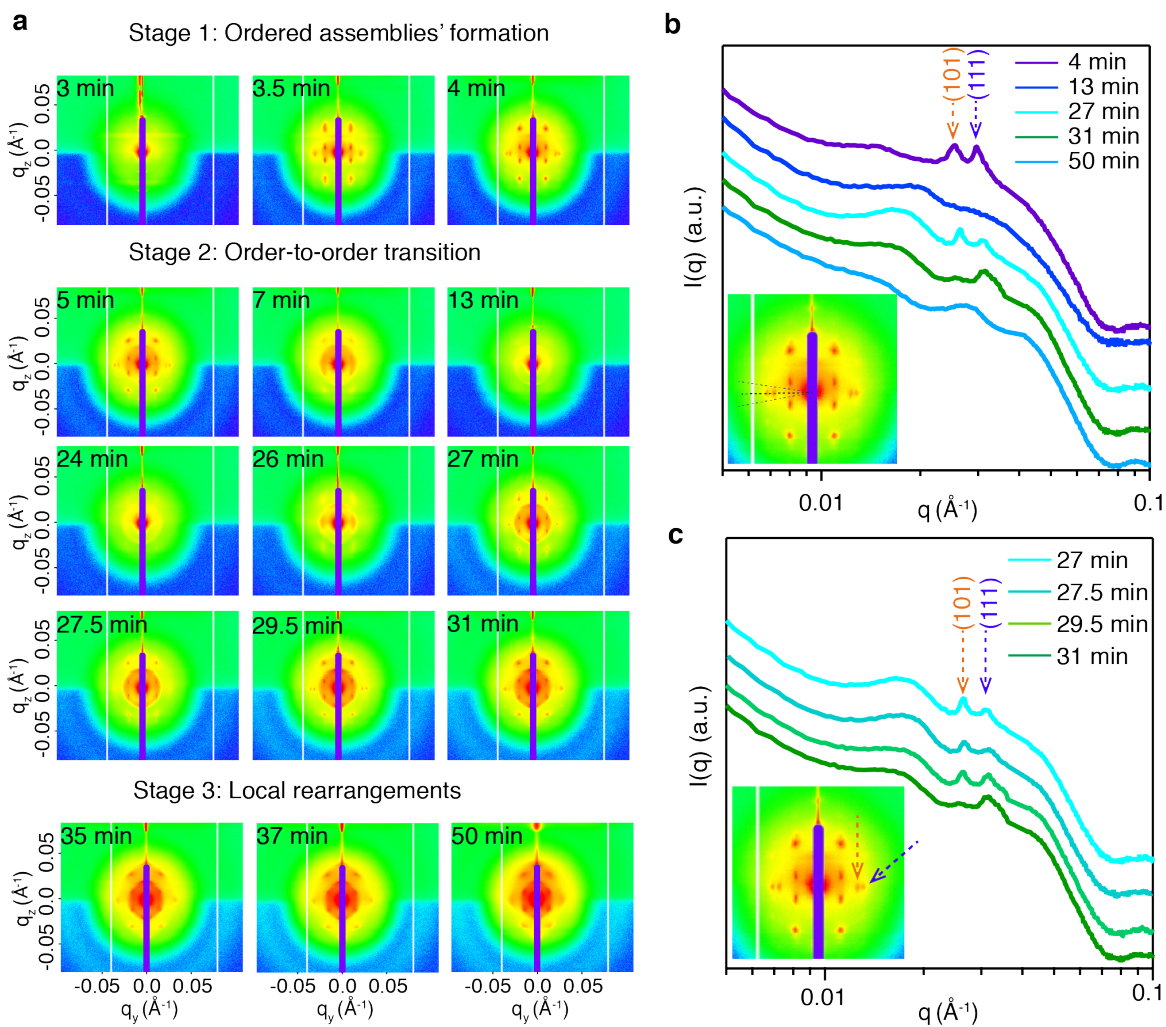

**Figure 4. The structural evolution of multicomponent nanocomposite during solvent evaporation. a,** the *in-situ* GTSAXS patterns at different stages during the drying process. **b, c,** the 1-D GTSAXS sector average profiles at different times. Inset in b shows the sector integrated region marked by black dash lines. The orange and blue arrows in the inset of c indicate the (101) of the "simple hexagon" structure and (111) of the "train track" structure, respectively.

**Conclusions**

In summary, the present study demonstrated a unique phase behavior of the multicomponent nanocomposites composed of NPs, BCP-based supramolecules, and small molecules, where the particle size is close to the polymer microdomain size. We observed a symmetry-breaking morphology named "train track" structure, which has never been achieved in polymer/NPs blends. The addition of NPs induces the polymer morphological transition from the cylinder to the perforated layer structure, and the asymmetric ordering of NPs stabilizes the metastable HPL morphology of BCP, leading to assembly beyond the native BCP morphology. However, its structure formation window in terms of the blend composition and NP size is narrow. Therefore, it is nontrivial to observe the structure experimentally, and it would provide insights into the future exploration of novel morphologies in BCP-based nanocomposites. We also observed another "simple hexagon" structure in the same blend, where polymer adopts a perforated layer structure with holes from adjacent layers aligned and NPs arrange into hexagonal lattice. The macroscopic phases in the same blend with different nanostructures are due to the self-sorting behavior based on several nanometer differences in NP size: smaller NPs form "train track" morphology while the larger NPs form "simple hexagon" morphology, indicating the strong self-regulation from different components in the multicomponent system based on the local need. The self-sorting also shows the possibility to control the microstructure based on NP size with high accuracy.

**Materials and Methods**

**Materials.** PS (33 kDa)-*b*-P4VP (8 kDa) (PDI = 1.10) was purchased from Polymer Source, Inc. 3-n-Pentadecylphenol (PDP) (90%−95%) was purchased from ACROS Organics. Chloroform was purchased from Fisher Scientific, and no HCl was detected using NMR. Iron oxide NP (15 nm) was purchased from Ocean Nanotech. No unexpected or unusually high safety hazards were encountered.

**Sample Preparation.** Supramolecule PS(33 kDa)-b-P4VP(8 kDa)(PDP)$_1$ was first dissolved in chloroform to form a 15 mg/ml solution. The desired amounts of small molecules were added to the solution and stirred overnight. NP suspensions were then mixed with solutions of supramolecule and small molecules. The ratio of the nanoparticle solution to the supramolecule/small molecule solution was controlled to reach the desired nanoparticle loading. The solution samples were directly used for solution scattering experiments. To prepare the bulk samples, ~ 300 μL of the blend solution was dried in a Teflon beaker at room temperature until ~40 μL (the solvent fraction is ~80%) of the solution remained. The solution was transferred to a small chamber, which caused the solvent to evaporate slowly, and dried overnight.

**Transmission Electron Microscopy (TEM).** Samples were embedded in resin (Araldite 502, Electron Microscopy Sciences) and cured at 60 °C overnight. Thin sections about 60 nm in thickness were microtomed using a RMC MT-X Ultramicrotome (Boeckler Instruments) and picked up on copper TEM grids on top of water. Samples were stained with iodine vapor to selectively stain the P4VP region. The thin sections were imaged using a FEI Tecnai 12 at the accelerating voltages of 120 kV.

**High-Angle Annular Dark-Field Scanning Transmission Electron Microscopy (HAADF-STEM) Tomography.** The projection images for 3D electron tomography was collected using a FEI TitanX 60-300 microscope with a 10 mrad probe semi-convergence angle operated at 200 kV at National Center for Electron Microscopy (NCEM) facility of the Molecular Foundry. The pixel size was 4.76 nm. A hummingbird heavy tomography holder was used to acquire a series of TEM images at tilt angles ranging ± 70° at an angular interval of 1°. The tilt series was aligned and reconstructed using the eTomo software of the IMOD tomography package. Reconstruction was done using the weighted-back-projection method. 3D visualization was performed using Tomviz. Slices through the reconstruction show the NPs as black dots, and white dots are common artifacts of the reconstruction process due to the missing information in the acquired data (the missing wedge).

**Small Angel X-ray Scattering (SAXS).** SAXS experiments were performed at Beamline 11-BM (Complex Material Scattering) of National Synchrotron Light Source-II (NSLS-II, Brookhaven National Laboratory). The SAXS data were collected on a Dectris 2M detector at a sample-to-detector distance of 2 m, using an X-ray beam with an energy of 13.5 keV (the corresponding wavelength $\lambda = 0.92$ Å). Images were plotted as intensity (I) vs q, where $q = (4\pi/\lambda) \sin(\theta)$, $\lambda$ is the wavelength of the incident X-ray beam, and $2\theta$ is the scattering angle. The circular-average profiles of SAXS patterns were extracted using Igor Pro with the Nika package. For SAXS structure fitting, we used CrystalMaker to build the structures and used CrystalDiffract to calculate the lattice diffraction patterns.

**Grazing Transmission Small-angle X-ray Scattering (GTSAXS).** GTSAXS experimental measurements were made at beamline 7.3.3 at the ALS in Lawrence Berkeley National with X-ray wavelength of 1.24 Å. The scattering intensity distribution was captured by a Pilatus 2 M detector. A 2 cm x 2 cm silicon substrate was placed in a chamber designed for *in situ* measurements and aligned with the beam. A 350μL chloroform reservoir was injected into the chamber to slow the drying process, and then ~100 μL of sample solution was drop cast onto the substrate. The measurements were taken at an incident angle of α=0.8° every 30 seconds. The sector-average profiles of SAXS patterns were extracted using Igor Pro with the Nika package.

**Dissipative Particle Dynamics (DPD) Simulation.** DPD simulation, a coarse-grain model in molecular dynamics was carried out to probe the interplay between each component and the spatial distribution of each component in blends. The simulation setup is based on the reparametrized DPD methods with simulation density being 5.0. There are four different types of soft particles in the system, which represent comb polymer (P4VP), coil polymer (PS), NP, and small molecule (PDP) respectively. The structures of supramolecules, PDP and nanoparticles are same with our previous study.[31] The beads in the same polymer are connected by harmonic bond:

$$F_{i,j}^{bond} = K(r_{ij} - r_0)\hat{r}_{ij} \tag{1}$$

Where i and j are interconnected beads, K is the spring constant and $r_0$ is the equilibrium distance. The freely joint chain model is used in the reparametrized DPD simulation, with K=50.0 and $r_0$=1.0.

Besides the harmonic bond, three non-bonded force fields exist between two random beads that are close to each other. They are the soft repulsion force, thermal fluctuation and drag force:

$$F_{ij}^C = \begin{cases} -a_{ij}(1-|r_{ij}|)\hat{r}_{ij} & if\ |r_{ij}| < 1 \\ 0 & if\ |r_{ij}| \geq 1 \end{cases} \tag{2}$$

$$F_{ij}^R = \sigma w^R(r_{ij})\theta_{ij}\hat{r}_{ij}\zeta/(\delta t)^{1/2} \tag{3}$$

$$F_{ij}^D = \frac{1}{2}\sigma^2\left(w^R(r_{ij})\right)^2/kT(v_{ij}\cdot\hat{r}_{ij})\hat{r}_{ij} \tag{4}$$

In Equations (2)-(4), $a_{ij}$ represents the maximum repulsion between particle type i and particle type j, $\zeta$ is a random variable with zero mean and variance one, and $w^R(r) = (1 - r)$ for $r < 1$ and $w^R = 0$ for $r > 1$. The $\sigma$ and $\delta t$, which are the noise factor and time step, take the value 0.10 and 0.015 respectively. The relationship between the soft repulsion parameter between two soft particles, $a_{ij}$, and the Flory-Huggins parameters are:

$$a_{ij} \approx a_{ii} + 1.45\chi_{ij} \qquad (5)$$

Where $a_{ii}$ characterizes the maximum repulsion between the same particle type and $a_{ii}$=15kT for DPD simulation density of 15.0. The detail parameter information is shown in Table S1.

In this case, the size of the simulation box is $44.2 \times 41.4 \times 38.5$. There are 96 nanoparticles, 7895 BCP chains and 189480 PDP particles (half are bonded to the BCP and half are free) in the simulation. The detail information on the structural setup of coil-comb polymer and NPs can be found in the previous paper on self-regulated coassembly.[23] The simulation consists of two steps: 1. the positions of the NPs are fixed as the "train track" structure, while the coil-comb BCP and small molecules can move freely based on the intermolecular interaction; 2. the positions of the NPs are unfixed, and all the four components of the system move based on the intermolecular interactions. When studying the order-to-order transition between "simple hexagon" and train track structure, the initial morphology in Step 1 is the "simple hexagon". Each simulation step has a relaxation period of 1,000,000 time steps.

**Acknowledgement:** This work was funded by the U.S. Department of Energy, Office of Science, Office of Basic Energy Sciences, Materials Sciences and Engineering Division under Contract No. DE-AC02-05-CH11231 (Organic-Inorganic Nanocomposites KC3104). Scattering studies were done at the Advanced Light Source and National Synchrotron Light Source-II at Brookhaven National Laboratory, were supported by the Office of Science, Office of Basic Energy Sciences, of the U.S. Department of Energy under Contract no. DE-AC02-05CH11231. Work performed at the Molecular Foundry was supported by the Office of Science, Office of Basic Energy Sciences, of the U.S. Department of Energy under Contract no. DE-AC02-05CH11231. We thank P. Bai and Y. Xiao for initial sample preparation and measurements.

**Author Contributions:** T.X. conceived the idea and guided the project. L.M. performed studies contributed to Figure 1, 2, and 4. H.H. and A.A. performed simulation in Figure 3. P.E. helped the tomography experiment.

**Notes:** The authors declare no competing financial interest.

**Supporting Information:**

The Supporting Information is available free of charge at XXX.

Supporting information includes TEM images, simulation results, SAXS fitting, GTSAXS profiles, interstitial size calculation, detailed NP size analysis, and relevant movies.

Supplementary Materials for

# Symmetry-Breaking and Self-Sorting in Block Copolymer-based Multicomponent Nanocomposites

Le Ma[1,2]†, Hejin Huang[3]†, Peter Ercius[4], Alfredo Alexander-Katz[3] and Ting Xu[1,2,5]*

[1]Department of Materials Science and Engineering, University of California, Berkeley, CA 94720, USA

[2]Materials Sciences Division, Lawrence Berkeley National Laboratory, Berkeley, CA 94720, USA

[3]Department of Materials Science and Engineering, Massachusetts Institute of Technology, Cambridge, MA, 02139, USA

[4]National Center for Electron Microscopy, Molecular Foundry, Lawrence Berkeley National Laboratory, Berkeley, CA 94720, USA

[5]Department of Chemistry, University of California, Berkeley, CA 94720, USA

† Authors contributed equally

*Corresponding author email: tingxu@berkeley.edu

**This PDF file includes:**

    Figures. S1 to S16
    NP size analysis
    Table S1
    Captions for Movies S1 to S5

**Other Supplementary Materials for this manuscript include the following:**

    Movies S1 to S5

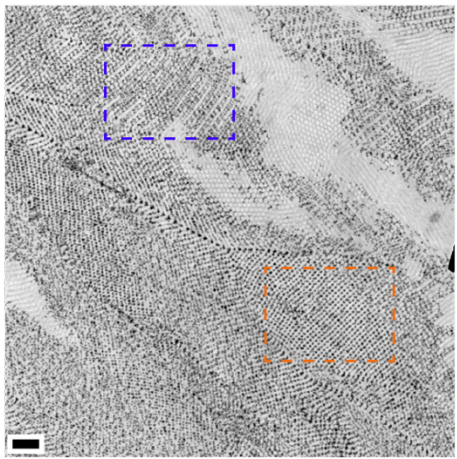

**Figure S1. TEM image of the nanocomposite shows two structures in the system.** One with a "train track" projection (blue box), the other with a "rectangular-like" lattice (orange box). Scale bar: 200 nm.

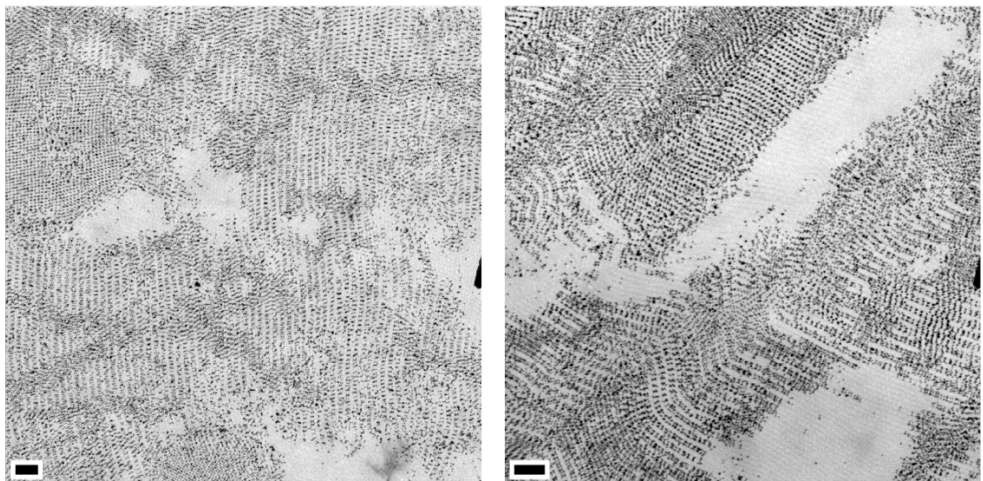

**Figure S2. TEM images of the "train track" structure after heating at 110°C overnight.** Scale bars: 200 nm.

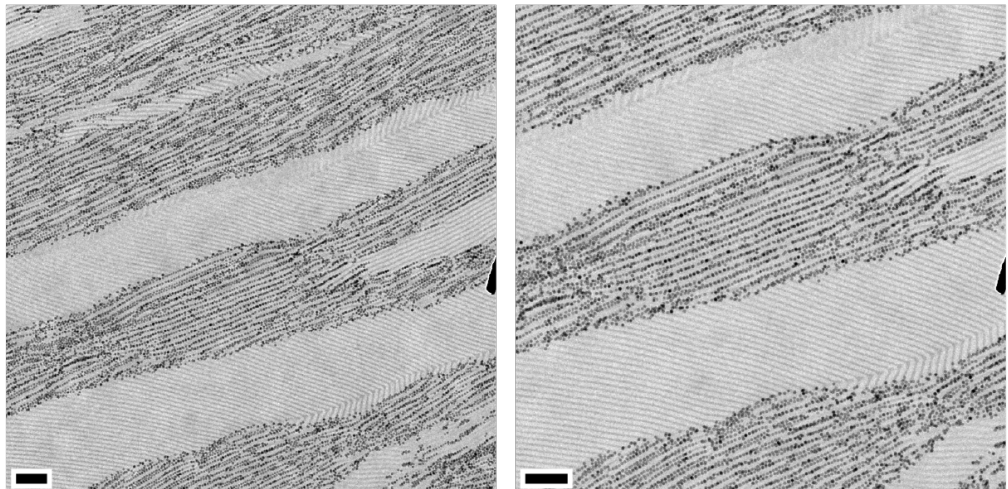

**Figure S3. TEM images of the morphology formed in supramolecule/15 nm blend without containing free small molecules.** NPs form chain-like morphology. Scale bars: 200 nm.

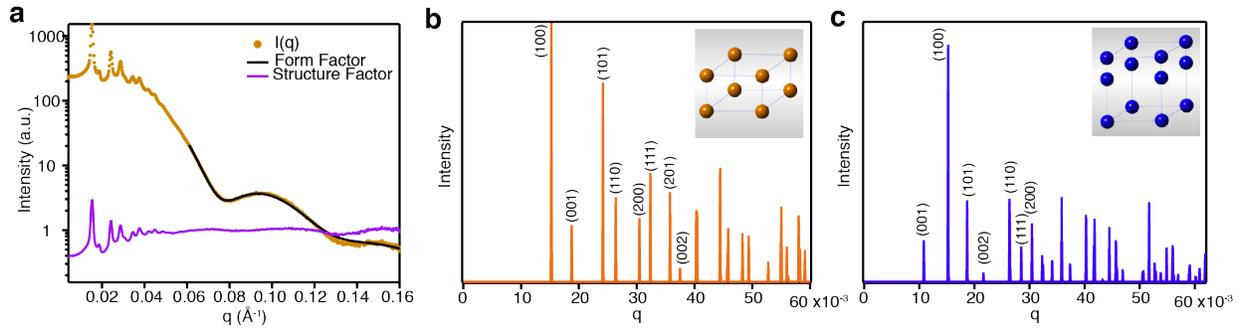

**Figure S4. SAXS results of the studied multicomponent polymer nanocomposite**. **a,** The form factor fitting and structure factor reduction. **b,** The calculated diffraction pattern for "simple hexagon" structure with lattice parameters: $a = b = 47.6\ nm,\ c = 33.5\ nm,\ \alpha = \beta = 90°,\ \gamma = 120°$. **c,** The calculated diffraction pattern for "train track" structure with lattice parameters: $a = b = 47.6\ nm,\ c = 58\ nm,\ \alpha = \beta = 90°,\ \gamma = 120°$. The inset shows the unit cell with another NP at ~1/3 of the c axis.

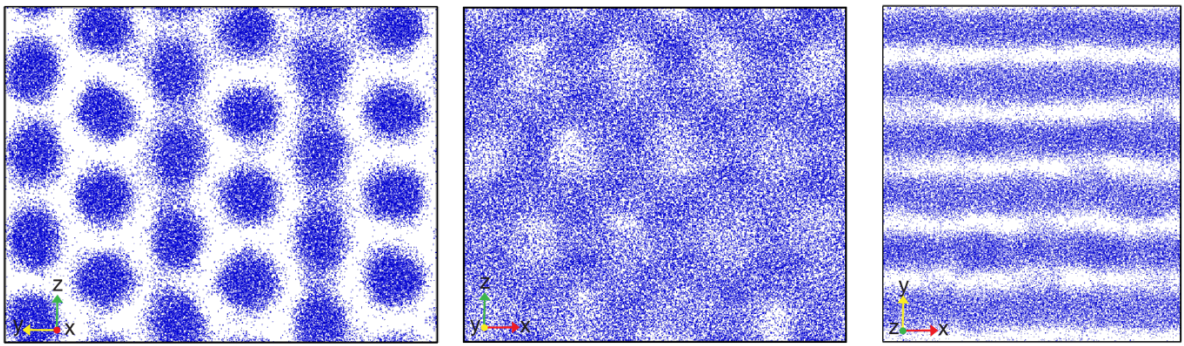

**Figure S5. Supramolecule adopts HPL morphology in the "train track" structure.** The DPD simulation results of the different projections of PS (coil-polymer) distribution in "train track" structure.

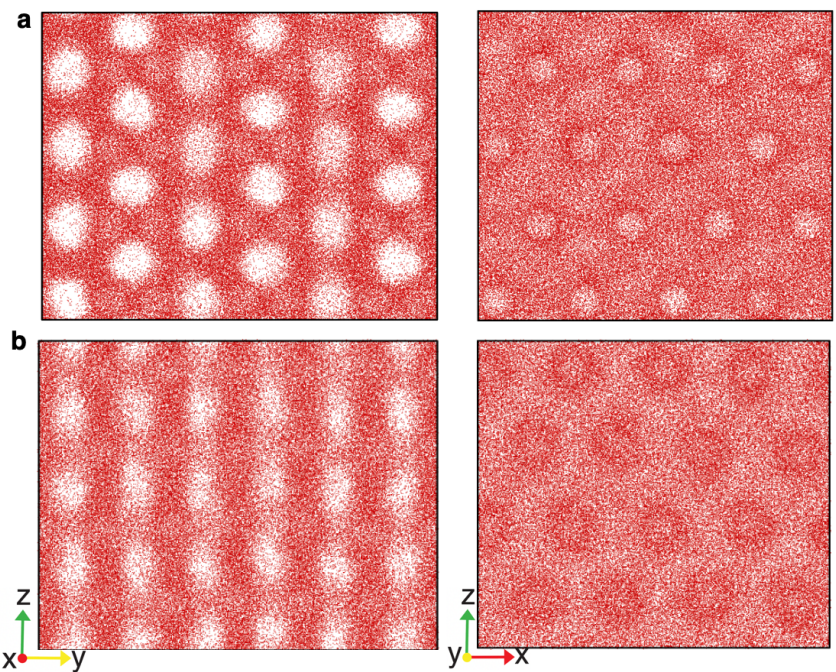

**Figure S6.** The DPD simulation results of the distribution of small molecules in the "train track" (a) and the "simple hexagon" structure (b).

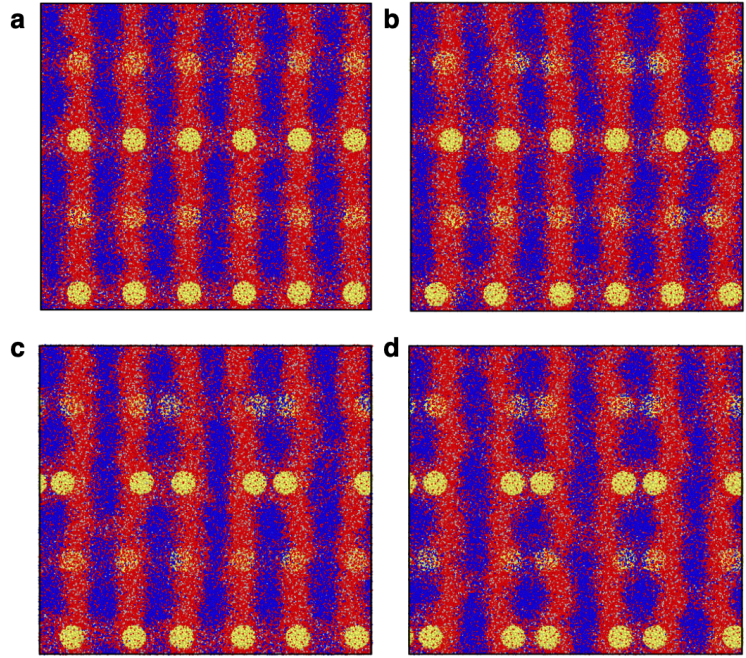

**Figure S7. The DPD simulation results of the "simple hexagon" to the "train track" transition.** Figure a is the initial state where particles are fixed in the "simple hexagon" morphology. Then we released the particle and captured their transition process steps. Each simulation step (a-d) has a relaxation period of 1,000,000 time steps. In step 1(Figure b), the first row of particles transitioned into the "train track" morphology. Then the second row changed to the "train track" structure after step 2 (Figure c). Gradually, all the particles finished the transition.

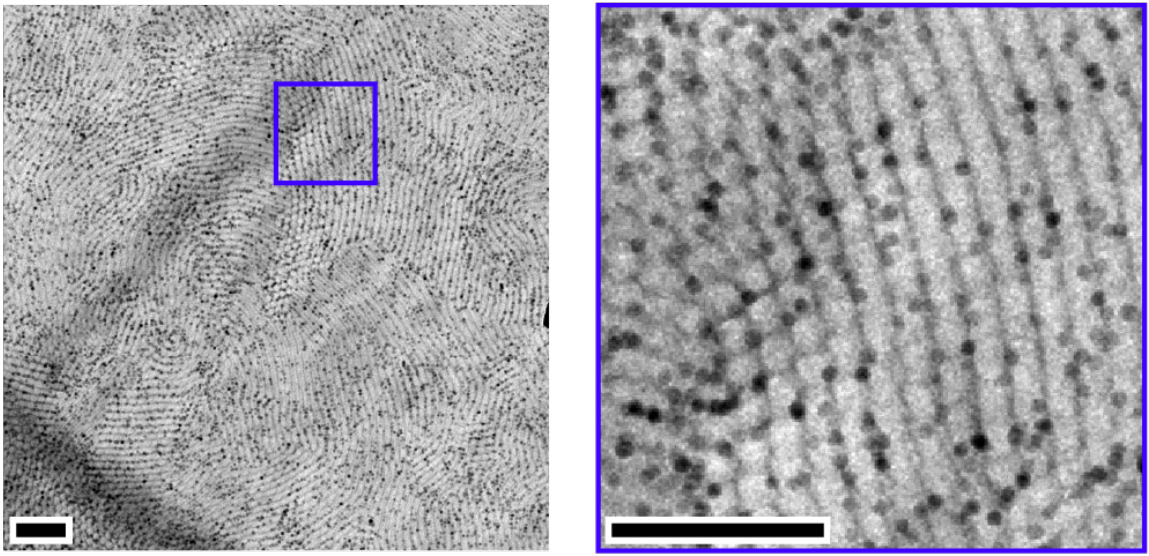

**Figure S8. TEM images of the multicomponent nanocomposite with fast evaporation rate.** The solution was dried within one hour from a solvent fraction of ~95%. We observed more "simple hexagon" morphology. Scale bars: 200 nm.

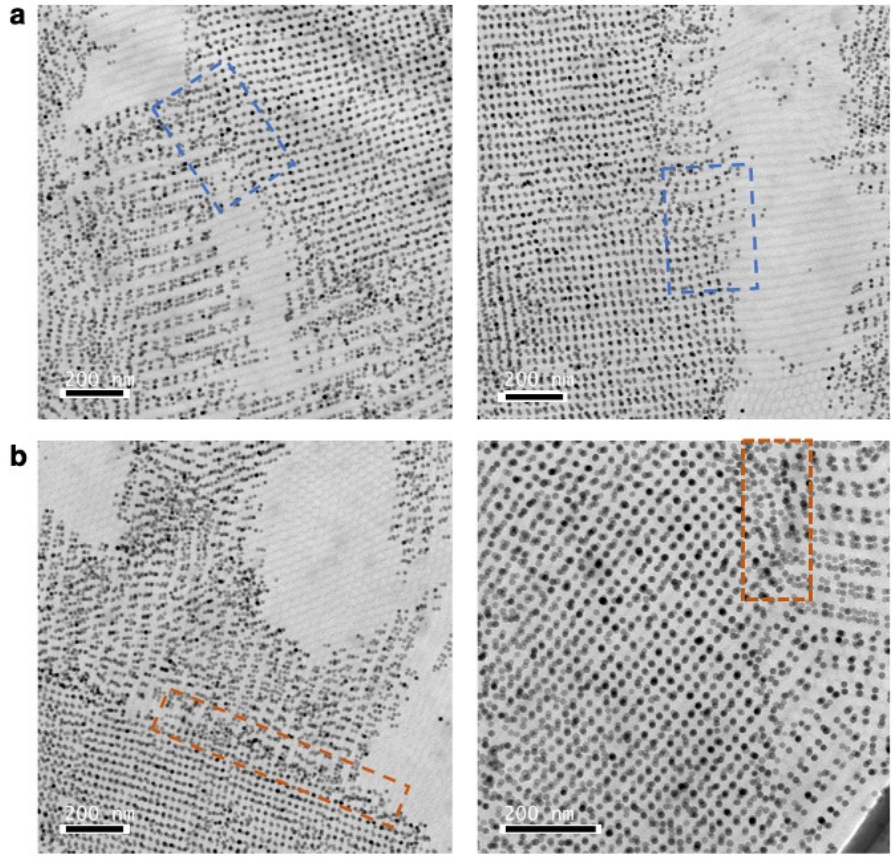

**Figure S9. TEM images of the interface between the "simple hexagon" and the "train track" morphologies. a,** The interface between the two morphologies where the lattice plane is aligned. We can see the continuous transition of two phases as indicated by the blue boxes. **b**, The interface between the two structures with different grain orientations. There is an apparent transition state between the two phases as marked by the orange boxes. Scale bars: 200 nm.

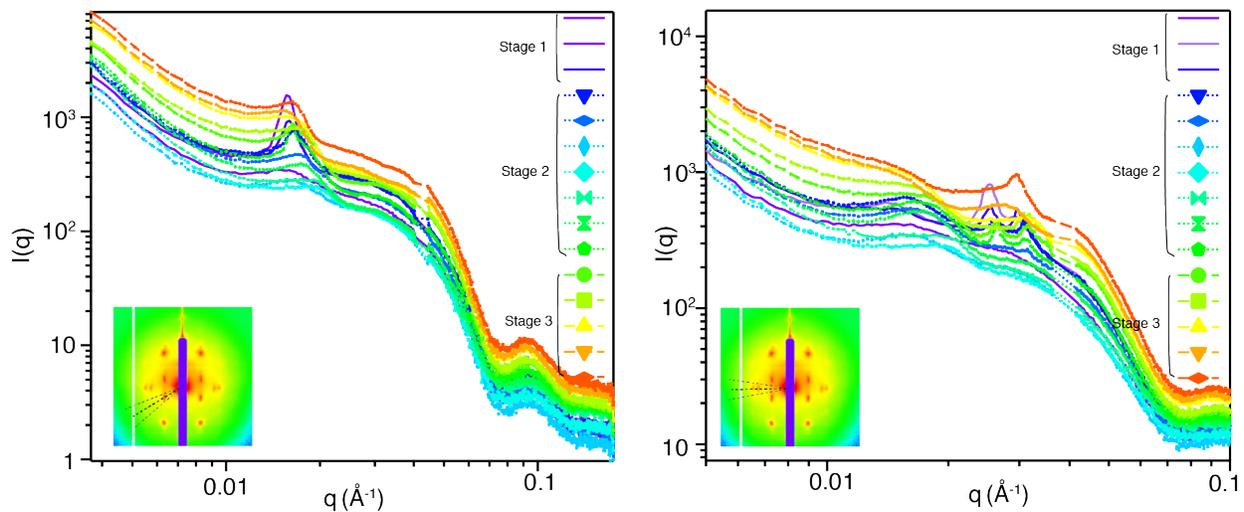

**Figure S10. 1-D GTSAXS profiles of the multicomponent polymer nanocomposite at different stages during the drying process.** Stage1: the ordered assemblies' formation; Stage 2: order-to-order transition; Stage 3: local rearrangement. The inset shows the sector averaged peaks.

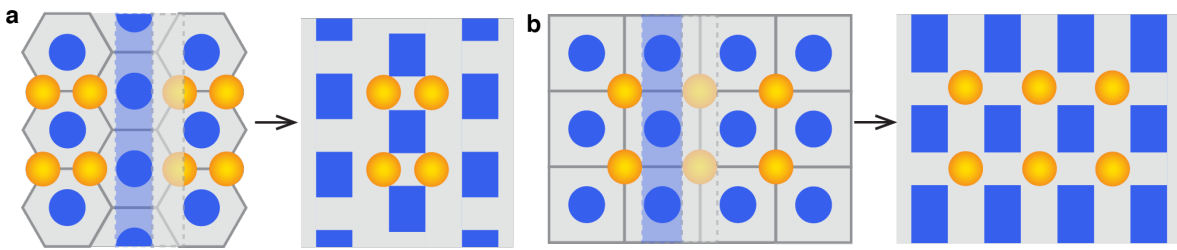

**Figure S11. NP induced morphology transition. a, b,** Schematic of the order-to-order transition: polymer transforms from cylinder to perforated layer structure in the "train track" structure (a) and the "simple hexagon" structure (b). The blue and gray blocks represent two polymer microdomains. Orange spheres represent NPs.

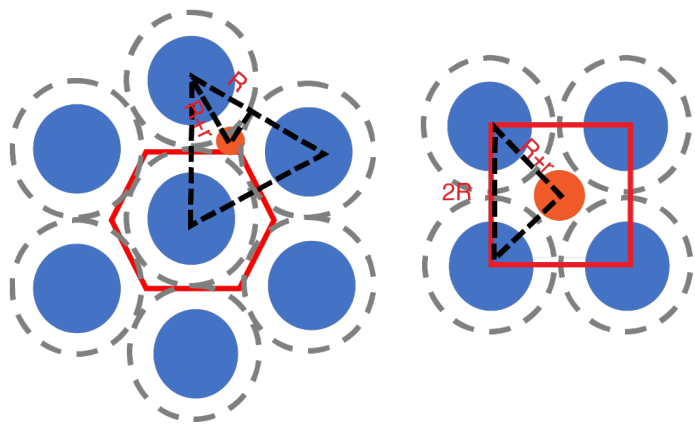

**Figure S12. Schematic representation of a cylindrical supramolecule morphology with hexagonally arranged cylinders and square-packed cylinders.** In a regular hexagonally packed cylindrical morphology, the interstitial site exists at the intersection of three adjacent cylinders. The radius of the interstitial site $r$, is related to the radius of the supramolecule cylinder, $R$, by $r = \left(\frac{2}{\sqrt{3}} - 1\right) R$. In the blend supramolecule/PDP(1), R = 18.9 nm. The calculated interstitial size is 5.9 nm. In the square-packed cylinders, $r = (\sqrt{2} - 1)R$, and the calculated interstitial size is 15.6 nm.

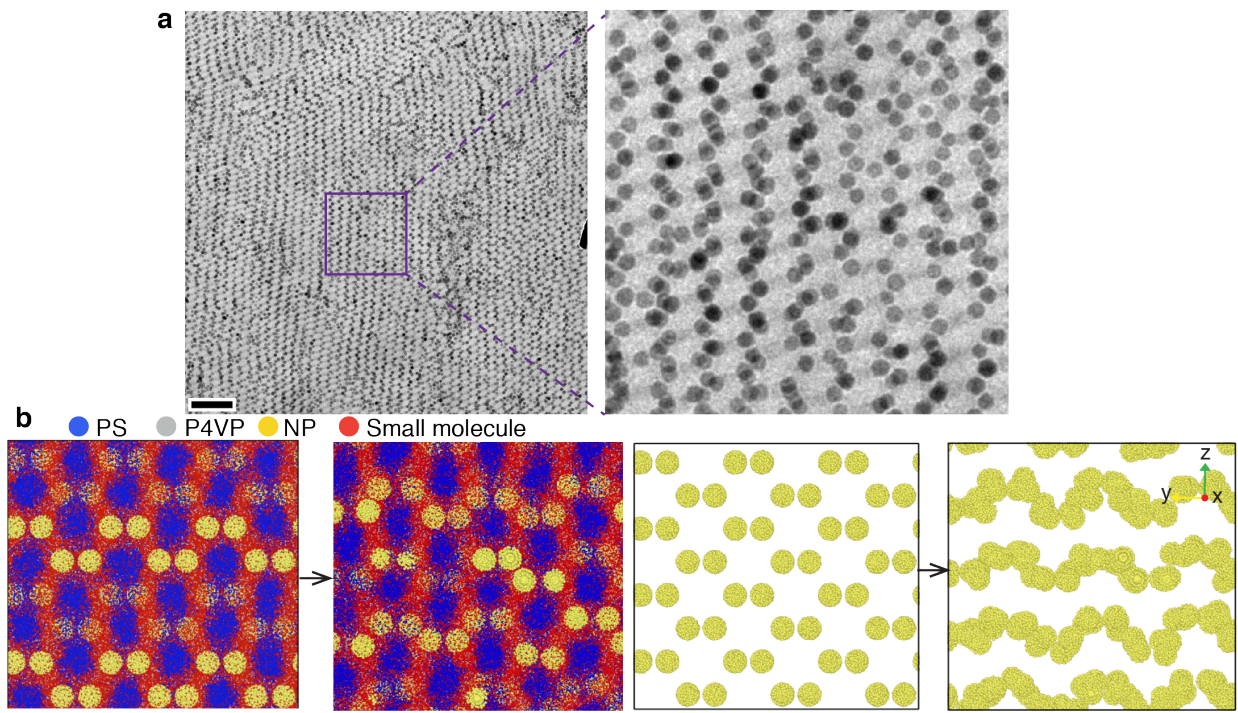

**Figure S13. The assembled morphology in the multicomponent system with high NP loading. a**, TEM images of the blend with doubled NP loading (10 vol%). Scale bars: 200 nm. **b**, The DPD simulation results for blends of coil-comb BCPs, small molecules, and NPs with doubled NP loading. The NPs are first fixed into the six corners of the hexagon as shown in left image, then their positions in the *y*-direction are released to allow oscillations of NPs. The right images show the morphologies after a relaxation period of one million time steps. The NPs cannot maintain the hexagonal morphology after relaxation, which indicates the large polymer chain deformation if the NPs occupy the six corners of the hexagon.

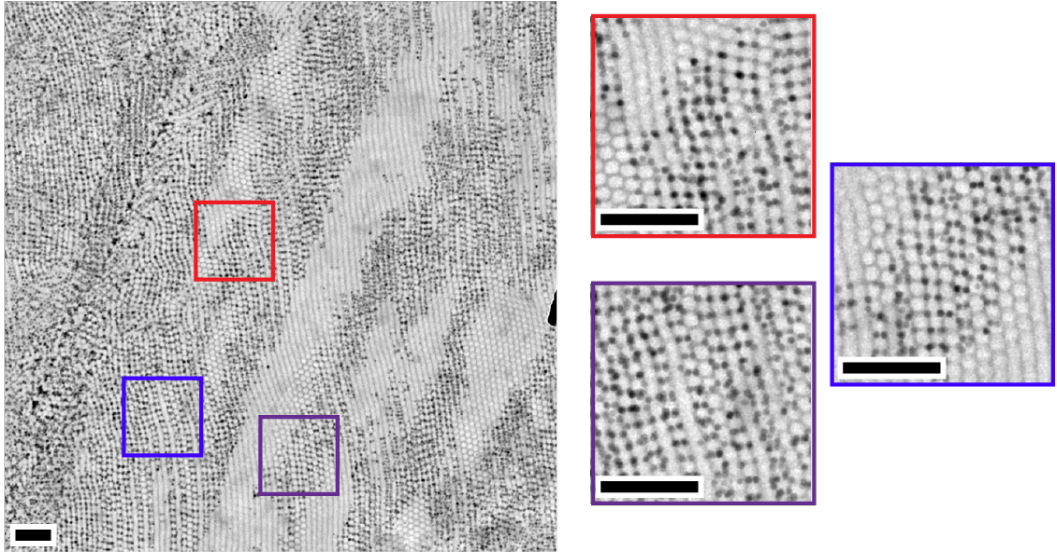

**Figure S14. TEM images of the blend that captures the intermediate state of structure evolution.** Scale bars: 200 nm.

**Nanoparticle size analysis**

1. <u>The nanoparticle size distribution analysis of the fresh purchased nanoparticle solution.</u>

We drop-cast the particle solution to a TEM grid and took the TEM image (Figure S13a). Then we used Image J to analyze the particle size distribution. The analysis procedure is shown in Figure S13. We first adjusted the threshold of the image to identify each individual particle. Then we had a binary image used for particle analysis (Figure S13b). Next, we used the menu command Analyze >Analyze particles; this will provide information about each particle in the image. Figure c is an overlay image of the analysis result (the resulting colored particles on top of the original particles), which provides us with the area information of each colored particle. Then we can extract the particle diameter information and plot their distribution. Figure S13d is the size distribution result of the 1289 particles.

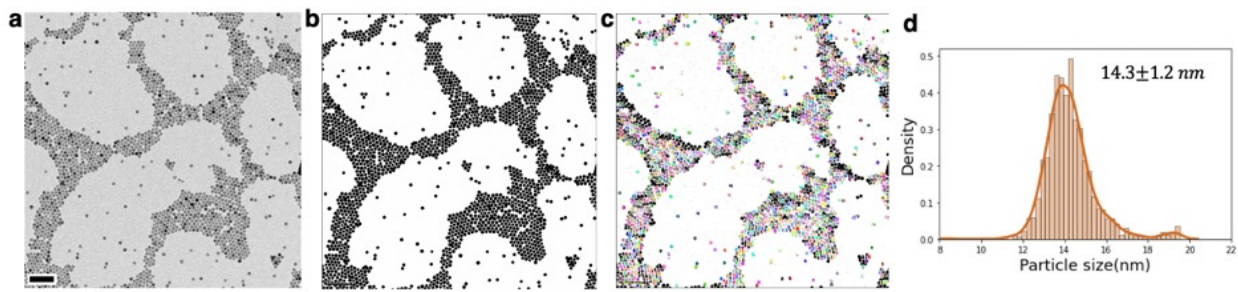

**Figure S15. TEM image and the size distribution of the NPs used in this study.** Scale bar: 100 nm.

2. <u>The nanoparticle size distribution analysis in the assembled morphologies.</u>

To investigate if there is any size difference in the "train track" and "simple hexagon" morphologies, we performed a similar analysis on the particles within these two structures separately. To make the statistical analysis, we chose different regions of "train track" and "simple hexagon" morphologies and selected particles more than 1000 to conduct the size distribution plot. We selected regions with projections along the *ac* or *bc* plane where particles from different layers are almost overlapped, and we can identify the individual particles, as shown in Figure S14. The size distribution result is shown in Figure 2b and 2c.

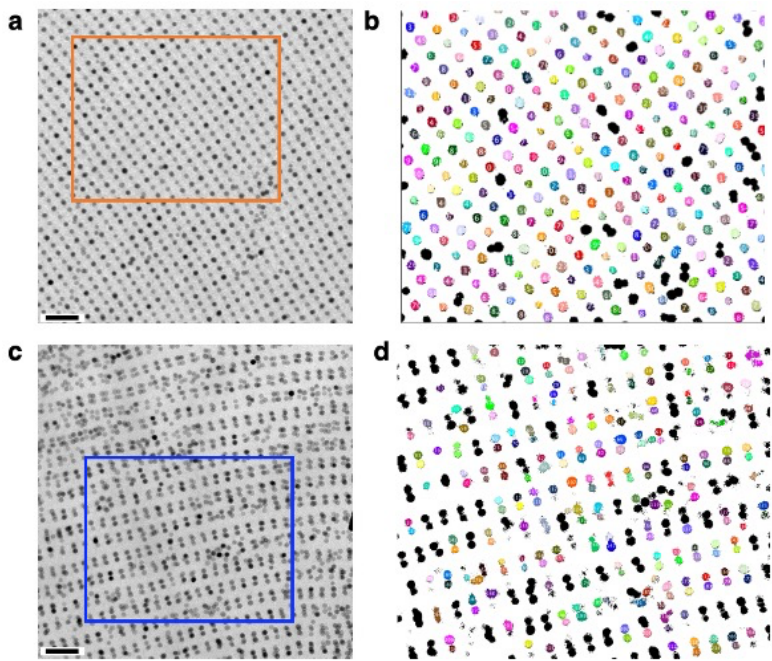

**Figure S16. TEM images and the size distribution analysis of the NPs in the "simple hexagon" (a, b) and the "train track" structures (c, d).** Figures b and d show the analysis results of the regions shown in boxes in Figures a and c. Scale bar: 100 nm. This figure shows an example analysis of one selected region. We selected several different areas of each morphology to perform the same analysis. We analyzed 1113 particles for the "train track" structure and 1102 particles for the "simple hexagon" structure.

**Table S1. The DPD simulation interaction parameters between different components**

| $a_{ij}$ | Interaction parameter |
|---|---|
| $a_{AB}$ | 34.6 |
| $a_{AC}$ | 15.0 |
| $a_{AD}$ | 10.0 |
| $a_{BC}$ | 17.9 |
| $a_{BD}$ | 16.4 |
| $a_{CC}$ | 15 |
| $a_{CD}$ | 5 |

A B C and D, which represent comb polymer (P4VP), coil polymer (PS), NP, and small molecule, respectively.

**Movie S1.** The tomography reconstruction result of NP 3-D lattice of the "train track" structure.

**Movie S2.** The tomography reconstruction result of NP 3-D lattice of the "simple hexagon" structure.

**Movie S3.** The DPD simulation results of NP arrangement after release from an initial "train track" lattice. blue: coil polymer (PS); gray: comb polymer(P4VP); yellow: NP; red: PDP.

**Movie S4.** The complete set of the GTSAXS patterns in the *in-situ* drying process of the multicomponent nanocomposite.

**Movie S5 (a, b).** The DPD simulation results of NP arrangement after release from an initial "simple hexagon" lattice. blue: coil polymer (PS); gray: comb polymer(P4VP); yellow: NP; red: PDP. **a**, $d/l \approx 1.0$, **b**, $d/l \approx 1.1$.